\documentstyle[12pt]{article}

\renewcommand{\baselinestretch}{1.05}

\begin{document}
\def\be{\begin{eqnarray}}
\def\en{\end{eqnarray}}
\newcommand{\ba}{\begin{array}}
\newcommand{\ea}{\end{array}}
\newcommand{\bd}{\begin{displaymath}}
\newcommand{\ed}{\end{displaymath}}
\newcommand{\beq}{\begin{equation}}
\newcommand{\eeq}{\end{equation}}
\def\ov{\overline}
\def\up{\uparrow}
\def\dw{\downarrow}
\def\non{\nonumber}
\def\la{\langle}
\def\ra{\rangle}
\def\nc{N_c^{\rm eff}}
\def\ep{\varepsilon}
\def\vp{\varepsilon}
\def\vma{{_{V-A}}}
\def\vpa{{_{V+A}}}
\def\m{\hat{m}}
\def\fp{{f_{\eta'}^{(\bar cc)}}}
\def\half{{{1\over 2}}}
\def\eq{eq.~}
\def\vcbud{{V_{cb}V_{ud}^*}}
\def\vcbcs{{V_{cb}V_{cs}^*}}

\def\bra{\langle}
\def\ket{\rangle}

\def\i{i}                  
\def\as{\mbox{$\alpha_s$}} 
\def\Oa{$O(\alpha_s)$}     
\def\Oaa{$O(\alpha_s^2)$}  
\def\msb{$\overline{ms}$~} 
\def\to{\rightarrow}       
\def\Im{Im}
\def\Re{Re}
\def\ln{\mbox{$\ell n$}}   
\def\sin{\mbox{ sin}}
\def\cos{\mbox{ cos}}
\def\Sum{\displaystyle\sum}
\def\eff{{\rm eff}}
\def\Heff{{\cal H}_{\rm eff}}
\def\ce{c^{\eff}}
\def\me{{\bf m}}
\def\r{{\bf r}}
\def\dsp{\displaystyle}
\def\nn{\nonumber}

\def\JPsi{J/\Psi}
\def\ov{\overline}
\def\etapp{{\eta^{(')}}}
\def\half{{1\over 2}}
\def\p{{(')}}
\def\pidone{{ K^{*-}  \rho^0}}
\def\pidtwo{{ \ov K^{*0}  \rho^-}}
\def\pidthree{{ K^{*-}  \omega}}
\def\pidfour{{ K^{*-}  \phi}}
\def\pidfive{{ K^{*-}  K^{*0}}}
\def\pidsix{{ \rho^{-}  \phi}}
\def\pidseven{{ \rho^{-}  \rho^0}}
\def\pideight{{\rho^{-}  \omega}}
\def\tmtwo{$\times10^{-2}$}
\def\tmthree{$\times10^{-3}$}
\def\tmfour{$\times10^{-4}$}
\def\tmfive{$\times10^{-5}$}
\def\tmsix{$\times10^{-6}$}
\def\tmseven{$\times10^{-7}$}
\def\tmeight{$\times10^{-8}$}
\def\tmnine{$\times10^{-9}$}
\def\tmten{$\times10^{-10}$}
\renewcommand{\arraystretch}{1.3}
\def\nc{\\}
\def\sc{\\[-1mm]}
\small

\def\empty{ (---) &&&&& }
\def\same{ }

\def\a{\alpha}
\def\b{\beta}
\def\g{\gamma}
\def\d{\delta}
\def\e{\epsilon}
\def\ve{\varepsilon}
\def\l{\lambda}
\def\m{\mu}
\def\n{\nu}
\def\G{\Gamma}
\def\D{\Delta}
\def\L{\Lambda}

\def\pr{{\sl Phys. Rev.}~}
\def\prl{{\sl Phys. Rev. Lett.}~}
\def\pl{{\sl Phys. Lett.}~}
\def\np{{\sl Nucl. Phys.}~}
\def\zp{{\sl Z. Phys.}~}
\font\el=cmbx10 scaled \magstep2 {\obeylines \hfill IP-ASTH-05-99
\hfill May 1999}

\vskip 1.5 cm

\centerline{\large\bf Charmless Hardronic Decays $B_u \to V V $:
Angular Distributions,} \centerline{\large\bf Direct CP Violation
and  Determination of the Unitary Triangle }
\medskip
\bigskip
\centerline{\bf ~ B. Tseng$^{a}$\footnote{E-mail:
btseng@phys.sinica.edu.tw} and Cheng-Wei Chiang
$^{b}$\footnote{E-mail: chengwei@andrew.cmu.edu} }
\bigskip
\centerline{$^a$ Institute of Physics, Academia Sinica}
\centerline{Taipei, Taiwan 115, Republic of China}
\bigskip
\centerline{$^b$ Department of Physics, Carnegie Mellon University}
\centerline{Pittsburgh, PA 15213, USA}
\bigskip

\centerline{\bf Abstract}
\bigskip
{\small Two-body charmless nonleptonic decays of $B_u \to V V$ are
studied within the generalized factorization approach using a
recent calculation of the effective Wilson coefficients $c^{\rm
eff}_i$, which are not only renormalization-scale and -scheme
independent but also gauge invariant and infrared finite. After
making a universal ansatz for the nonfactorizable contributions,
we parametrize these effects in terms of $N_c^{\rm eff}(LL)$ and
$N_c^{\rm eff}(LR)$, the effective numbers of colors arising from
$(V-A)(V-A)$ and $(V-A)(V+A)$ four-quark operators, respectively.
Three different schemes for these contributions are considered:
(i) the naive factorization, (ii) the large-$N_c$ improved
factorization, and (iii) our preferred choice: $(N_c^{\rm
eff}(LL),N_c^{\rm eff}(RR))=(2,5)$. We present the full angular
distribution of all charmless $B_u \to V V$ decays in both
 transversity  and  helicity frames.
Direct CP violation in these normalized angular correlation
coefficients is not negligible in $B^-_u \to K^{*-} \rho^0, K^{*-}
\omega$, and direct CP violation in the partial rate difference
for  $B^-_u \to K^{*-} \omega, K^{*-} \rho$ and $\rho^- \omega$
can be as large as
 $45\%$,  $25\%$,  $-10\%$, respectively. Due to the sizable QCD
penguin contributions in $\rho^- \omega$, the determination of the
unitary triangle $\alpha$ via this decay mode is more promising
than via $\rho^- \rho^0$. It is also encouraging to determine the
unitary triangle $\gamma$ through
 $B^-_u \to K^{*-} \rho$  because of $N_c$-insensitivity and
 the not-so-small tree contribution.
The impacts of a negative $\rho$ on the branching ratios and  CP
violation  are studied. We also comment on the theoretical
uncertainties and their possible impacts.
 } \vfill
\pagebreak

{\bf 1.}~~In past years we have witnessed remarkable progress in
the study of exclusive charmless $B$ decays. Experimentally, CLEO
\cite{CLEO,GW} has discovered many new two-body decay modes
\be
B\to\eta' K^\pm,~\eta' K^0,~\pi^\pm K^0,~\pi^\pm K^\mp,~\pi^0
K^\pm,~\pi^\pm K^{*\mp},~\rho^0\pi^\pm,~\rho^\mp \pi^\pm,~\omega
K^\pm,
\en
and found a possible evidence for $B\to \phi K^*$. Moreover, CLEO
has provided new improved upper limits for many other decay modes.
While all the measured channels are penguin dominated, the most
recently measured $\rho^0\pi^-$ and $ \rho^\mp \pi^\pm$  modes are
dominated by the tree diagrams. In the meantime, updates and new
results of many $B\to PV$ decays with $P=\eta,\eta',\pi,K$ and
$V=\omega,\phi,\rho, K^*$ as well as $B\to PP$ decays will be
available soon. With the $B$ factories Babar and Belle starting to
collect data, many exciting and harvest years in the arena of $B$
physics and $CP$ violation are expected to come.

An earlier systematic study of exclusive nonleptonic two-body
decays of $B$ mesons was made in \cite{Chau1}. Since then, many
significant improvements and developments have been achieved over
past years \cite{DEV}. For example, a next-to-leading order
effective Hamiltonian \cite{Buras92, Ciuchini, Buras96} for
current-current operators and QCD as well as electroweak penguin
operators \cite{EW} becomes available. The renormalization scheme
and scale problems with the factorization approach for matrix
elements can be circumvented by employing scale- and
scheme-independent effective Wilson coefficients. Heavy-to-light
form factors have been computed using QCD sum rules, lattice QCD
and potential models. Besides, the gauge and infrared regulator
dependence problem  of the effective Wilson coefficients has also
been resolved in \cite{CLY}. Finally, a theoretical framework,
namely the generalized factroization approach \cite{Cheng}, has
been shown to be  useful for the understanding of the experimental
data.

In our previous studies \cite{CT97,CCTK99,BS}, we have completed
all  branching ratios of  $B_{u,d,s} \to PP , VP$ and $ VV$. It is
known that there is rich physics in the $ V V$ decay modes
\cite{Valencia89,KPS}: in addition to the average quantity such as
the branching ratio, there are more observables in the VV modes
shown in the angular distribution, from which we can get more
information on the dynamics. Aside from the partial rate
difference, there are also more CP-violating observables in the
angular distributions of  $VV$ decay modes which can provide more
contents with  possible fingerprints of new physics beyond the
standard model. The advantage of having a large number of
observables in $ VV$ modes also results in some useful strategies
of determining the unitary triangles \cite{AS}. While the
contribution of the electroweak penguin (EWP), which does show
effects on the $\rho^- \rho^0$ decay mode \cite{KL}, is neglected
in some early studies \cite{KPS}, we include it in our
calculations. Meanwhile, all these earlier studies \cite{KPS, KL}
suffer from the gauge and infrared regulator dependence problem of
the effective Wilson coefficients. In this letter, we present an
updated analysis based on the gauge invariant effective Wilson
coefficients. The helicity  and transversity amplitudes appearing
in the angular distributions of all $B_u \to VV$ and their CP
violating observables are calculated. Topics about determination
of the possible CP-conserving final state interaction (FSI) phases
and/or possible CP-violating phases from  new physics, a revival
possibility of negative $\rho$ and its impact on  CP violation and
the determination of the unitary triangle are discussed in brief.

\medskip
{\bf 2.}~~Let us begin with a brief description of the theoretical
framework. The relevant effective $\Delta B=1$ weak Hamiltonian is
\be
{\cal H}_{\rm eff}(\Delta B=1) = {G_F\over\sqrt{2}}\Big[ V_{ub}V_{uq}^*(c_1
O_1^u+c_2O_2^u)+V_{cb}V_{cq}^*(c_1O_1^c+c_2O_2^c)
-V_{tb}V_{tq}^*\sum^{10}_{i=3}c_iO_i\Big]+{\rm h.c.},
\en
where $q=d,s$, and
\be
&& O_1^u= (\bar ub)_\vma(\bar qu)_\vma, \qquad\qquad\quad~~O_1^c =
(\bar cb)_ \vma(\bar qc)_\vma,   \non \\ && O_2^u = (\bar
qb)_\vma(\bar uu)_\vma, \qquad \qquad \quad~~O_2^c = (\bar
qb)_\vma(\bar cc)_\vma,   \non \\ && O_{3(5)}=(\bar
qb)_\vma\sum_{q'}(\bar q'q')_{\vma(\vpa)},  \qquad O_{4(6)}=(\bar
q_ \alpha b_\beta)_\vma\sum_{q'}(\bar q'_\beta
q'_\alpha)_{\vma(\vpa)}, \non \\ && O_{7(9)}={3\over 2}(\bar
qb)_\vma\sum_{q'}e_{q'}(\bar q'q')_{\vpa(\vma)}, \quad
O_{8(10)}={3\over 2}(\bar q_\alpha
b_\beta)_\vma\sum_{q'}e_{q'}(\bar q'_\beta
q'_\alpha)_{\vpa(\vma)},
\en
with $(\bar q_1q_2)_{_{V\pm A}}\equiv\bar q_1\gamma_\mu(1\pm
\gamma_5)q_2$. In Eq.~(2), $O_{3-6}$ are QCD penguin operators and
$O_{7-10}$ are electroweak penguin operators, and $c_i(\mu)$ are
Wilson coefficients which have been evaluated to the
next-to-leading order (NLO) \cite{Buras92,Ciuchini}. One important
feature of the NLO calculation is the renormalization-scheme
dependence of the Wilson coefficients (for a review, see
\cite{Buras96}). In order to ensure the $\mu$ and renormalization
scheme independence for the physical amplitude, the matrix
elements, which are evaluated under the factorization hypothesis,
have to be computed in the same renormalization scheme and
renormalized at the same scale as $c_i(\mu)$. However, as
emphasized in \cite{Cheng},  the matrix element $\la O\ra_{\rm
fact}$ is scale independent under the factorization approach and
hence it cannot be identified with $\la O(\mu)\ra$. Incorporating
QCD and electroweak corrections to the four-quark operators, we
can redefine $c_i(\mu)\la O_i(\mu)\ra= {c}_i^{\rm eff}\la
O_i\ra_{\rm tree}$, so that ${c}_i^{\rm eff}$ are  renormaliztion
scheme and scale independent. Then the factorization approximation
is applied to the hadronic matrix elements of the operator $O$ at
tree level. Recently, the controversy on gauge dependence and
infrared singularity associated with the effective Wilson
coefficients, criticized in \cite{Buras9806}, is resolved in
\cite{CLY}: Gauge invariance of the decay amplitude is maintained
under radiative corrections by assuming on-shell external quarks.
(for a more detailed discussion, see \cite{CLY, CCTK99}). In this
letter, we will utilize these recently-calculated gauge-invariant
Wilson coefficients  and thus our results do not suffer from these
gauge dependence and infrared singularity controversies. The
numerical values for ${c}_i^{\rm eff}$ are shown in  the last
column of Table I of \cite{CCTK99}, where $\mu={m}_b(m_b)$,
$\Lambda^{(5)}_{\overline{\rm MS}}=225$ MeV, $m_t=170$ GeV and
$k^2=m_b^2/2$ are used. From which we can see that the Wilson
coefficients for $b \to s$ and $\bar b \to \bar s$ are almost the
same and those for $b \to d$ and $\bar b \to \bar d$ are slightly
different.

In the naive factorization approach, only the factorizable
contributions are considered. However, as indicated by the $B \to
D P(V)$, contributions from the nonfactorizable amplitudes, which
cannot be calculated in this naive factorization approach, are
important for the understanding of the data \cite{NS,CT95}. The
spirit of the generalized factorization approach is to incorporate
these nonfactorizable contributions in a phenomenological way: we
parametrize these contributions, determine them from a few decay
modes and then make predictions for  the other modes. For the
$B\,(D)\to PP,~PV$ decays ($P$: pseudoscalar meson, $V$: vector
meson), there is only one single form factor (or Lorentz scalar)
involved in the decay amplitude. Thus, the effects of
nonfactorization can be lumped into the effective parameters
$a_i^{\rm eff}$ \cite{Cheng, Kamal}:
\be
a_{2i}^{\rm eff}=c_{2i}^{\rm eff}+c_{2i-1}^{\rm eff}\left({1\over
N_c}+\chi_{2i} \right),\qquad
a_{2i-1}^{\rm eff}=c_{2i-1}^{\rm eff}+c_{2i}^{\rm eff}\left({1\over
N_c}+\chi_{2i-1}\right),
\en
where $c_{2i,2i-1}^{\rm eff}$ are the Wilson coefficients of the
4-quark operators, and nonfactorizable contributions are
characterized by the parameters $\chi_{2i}$ and $\chi_{2i-1}$. We
can parametrize the nonfactorizable contributions by defining an
effective number of colors $N_c^{\rm eff}$, called $1/\xi$ in
\cite{BSW}, as $ 1/N_c^{\rm eff} \equiv (1/N_c)+\chi$. Thus the
nonfactorizable effects are effectively incorporated in the
factorization approach, that is the generalized factorization
framework.
However, the general amplitude of $B(D)\to VV$ decay consists of
three independent Lorentz scalars, corresponding to the $S$-, $P$-
and $D$-wave amplitudes. Consequently, it is in general not
possible to define effective $a_i$ unless nonfactorizable terms
contribute in equal weight to all covariant amplitudes. In this
letter, we, following \cite{CCTK99,KPS,KL,AKL}, make a further
universal assumption for the nonfactorizable contributions to the
different invariant amplitudes, {\it i.e.}
$\chi_{A1}=\chi_{A2}=\chi_{A3}=\chi_{V}$, so that all the
nonfactorizable contributions for the different covariant
amplitudes are equally weighted.

Different factorization appraoches used in the literature can be
classified by the effective number of colors $N_c^{\rm eff}$. The
so-called ``naive" factorization discards all the nonfactorizable
contributions and takes $ 1/N_c^{\rm eff}= 1/N_c=1/3 $, whereas
the ``large-$N_c$ improved" factorization \cite{Buras} drops out
all the subleading $1/N_c$ terms and takes $ 1/N_c^{\rm eff}=0$.
In this paper, in addition to predictions from these two
``homogeneous" nonfactorizable pictures, which assume that
$(N_c^{\rm eff})_1 \approx (N_c^{\rm eff})_2 \approx \cdots
\approx (N_c^{\rm eff})_{10}$, we also present results from the
``heterogeneous" one, which considers the possibility of $N_c^{\rm
eff}(LR)\neq N_c^{\rm eff}(LL)$. The consideration of the
``homogeneous" nonfactorizable contributions, which is commonly
used in the literature, has its advantage of simplicity. However,
as argued in \cite{CT97}, due to the different Dirac structure of
the Fierz transformation, nonfactorizable effects in the matrix
elements of $(V-A)(V+A)$ operators are {\it a priori} different
from that of $(V-A)(V-A)$ operators, i.e. $\chi(LR)\neq \chi(LL)$.
Since $1/N_c^{\rm eff}=1/N_c+\chi$ , theoretically it is expected
that
\be
&& N_c^{\rm eff}(LL)\equiv \left(N_c^{\rm
eff}\right)_1\approx\left(N_c^{\rm eff}\right)_2\approx
\left(N_c^{\rm eff}\right)_3\approx\left(N_c^{\rm
eff}\right)_4\approx \left(N_c^{\rm eff}\right)_9\approx
\left(N_c^{\rm eff}\right)_{10},   \non\\ && N_c^{\rm
eff}(LR)\equiv \left(N_c^{\rm eff}\right)_5\approx\left(N_c^{\rm
eff}\right)_6\approx \left(N_c^{\rm eff}\right)_7\approx
\left(N_c^{\rm eff}\right)_8 .
\en

We can thus make predictions based on different schemes for these
nonfactorizable contributions as done in \cite{CCTK99}. The  main
goal of this studies is to make predictions as much as possible
with effective one set indicated by the limited experimental data.
This ``minimal-fitting and global-predictions" will make
theoretical analysis simple and powerful. In this short letter, we
will use three different sets: (i) the naive factorization, (ii)
the large-$N_c$ improved factorization, and (iii) our the
preferred choice $(N_c(LL),N_c(LR) \approx (2,5)$. The first two
schemes are used as a reference  and the third set is based on our
analysis of the recent experimental data from CLEO (readers are
referred to \cite{CCTK99}).

Let's briefly discuss the input parameters: four Wolfenstein
parameters characterizing the Cabibbo-Kobayashi-Maskawa (CKM)
matrix are used with $\lambda=0.2205$ and $A=0.815$. As for the
parameters $\rho$ and $\eta$, different updated analyses
\cite{Parodi,Mele, AliLondon} have been performed. In these fits,
it is clear that $\sqrt{\rho^2+\eta^2}=0.41$ is slightly larger
than the previous analysis. For our purposes in the present paper
we will employ the values $\rho=0.175$ and $\eta=0.370$. Though
not be completely ruled out, a negative $\rho$ is disfavored by
these global analyses. However, it is found that a negative $\rho$
is preferable by the CLEO data \cite{Hou98, CCTK99}. The impact of
a negative $\rho$ is discussed with a simple sign-flip of $\rho$.
 Under the factorization hypothesis, the decay
amplitudes are expressed as the products of  decay constants and
form factors. We follow the standard parameterizations for decay
constants and form factors \cite{BSW}. For values of the decay
constants, we take $f_\pi=132$ MeV, $f_ K=160$ MeV, $f_\rho=210$
MeV, $f_{K^*}=221$ MeV,  $f_ \omega=195$ MeV and $f_\phi=237$ MeV.
Concerning the  heavy-to-light mesonic form factors, we will use
the BSW results  evaluated in the relativistic quark model
\cite{BSW} with the proper $q^2$-dependence adopted from the
heavy-quark symmetry.

\medskip
{\bf 3.}~~To set up our notation, we will briefly discuss the
angular distributions of $ B \to VV$ and  CP violating
observables.  The most general covariant amplitude for a $B$ meson
decaying into a pair of vector mesons has the form
\cite{Valencia89, KPS}:
\begin{equation}
A(\displaystyle B(p)\to V_1(k) V_2(q))=
\displaystyle\epsilon_{V_1}^{*\mu}\epsilon_{V_2}^{*\nu} \left( a
g_{\mu\nu}
+ \frac{b}{m_{V_1} m_{V_2}} p_{\mu} p_{\nu}
+ i \frac{c}{m_{V_1} m_{V_2}} \epsilon_{\mu\nu\alpha\beta}
k^\alpha q^\beta \right),
\end{equation}
where, $\epsilon_{V_1}$, $\epsilon_{V_2}$ and $m_{V_1}$, $m_{V_2}$
represent the polarization vectors and masses of the vector mesons
$V_1$ and $V_2$, respectively.  These invariant amplitudes $a$, $b$,
and $c$ have the advantage of being directly related to the decay
constants and form factors under the generalized factorization
approach.

However, it is customary to express the angular distributions of
$B \to VV$, with each vector meson subsequently decaying into two
particles, in terms of the helicity amplitudes, for which we use
the notation:
$H_{\lambda} = \left< V_1(\lambda) V_2(\lambda) \right| {\mathcal
H}_{wk} \left| B \right>$
%
for $\lambda = 0, \pm 1$.  The relations between the helicity and
invariant amplitudes are $H_0 = -a \, x - b \left( x^2 - 1
\right)$, and $H_{\pm} = a \pm \sqrt{x^2 - 1} \, c$. In general,
the explicit form of the angular distribution depends upon the
spin of the decay products of the two decaying vector mesons. To
be specific, we will take for the purpose of demonstration the
angular distribution of the decays $B \to V_1 (\to P_1
P_1^{\prime}) \, V_2 (\to P_2 P_2^{\prime})$, where
$P_1^{({\prime})}$ and $P_2^{({\prime})}$ denote pseudoscalar
mesons.  An example is $B^- \to K^{*-} \rho^0 \rightarrow
\left(K\pi\right)^-\left(\pi^+\pi^-\right)$. The normalized
angular distribution for this type of decay is:
\be
\label{distribution} \frac{1}{\Gamma}
\frac{d^3\Gamma}{d\cos\theta_1 d\cos\theta_2 d\phi} & = &
{\frac{9}{8 \pi}} \biggl \{ \frac{1}{4} \frac{\Gamma_T}{\Gamma} \,
\sin^2\theta_1 \sin^2\theta_2 + \frac{\Gamma_L}{\Gamma} \,
\cos^2\theta_1 \cos^2\theta_2 \nonumber \\ && + \frac{1}{4}\sin
2\theta_1 \sin 2\theta_2 \left[ \alpha_1 \, \cos\phi - \beta_1 \,
\sin\phi \right] \\ && + \frac{1}{2}\sin^2\theta_1 \sin^2\theta_2
\left[ \alpha_2 \cos2\phi - \beta_2 \sin2\phi \right] \biggr \},
\nonumber \en
where \be\ba{llllll} \frac{\Gamma_T}{\Gamma} & = & \frac{\vert
H_{+1}\vert ^2 + \vert H_{-1}\vert ^2} {\vert H_0\vert ^2 + \vert
H_{+1}\vert ^2 + \vert H_{-1}\vert ^2}, \ \ \ \ \ &
\frac{\Gamma_L}{\Gamma} & = & \frac{\vert H_0\vert ^2}{\vert
H_0\vert ^2 + \vert H_{+1}\vert ^2 +\vert H_{-1} \vert ^2},
\\[5mm]
\alpha_1 & = & \frac{Re\left (H_{+1}H_0^\ast +
H_{-1}H_0^\ast\right )} {\vert H_0\vert ^2 + \vert H_{+1}\vert ^2
+ \vert H_{-1}\vert ^2}, & \beta_1 & = &\frac{Im\left (H_{+1}
H_0^\ast- H_{-1}H_0^\ast\right)} {\vert H_0 \vert ^2+ \vert H_{+1}
\vert ^2 + \vert H_{-1} \vert ^2 },\\[5mm]
\alpha_2 & = & \frac{Re\left(H_{+1}H_{-1}^\ast\right)} {\vert
H_0\vert ^2 + \vert   H_{+1}\vert ^2 + \vert H_{-1}\vert ^2}, &
\beta_2 & = & \frac{Im\left (H_{+1} H_{-1}^\ast\right)} {\vert
H_0\vert ^2 +\vert H_{+1}\vert ^2 + \vert H_{-1}\vert ^2}. \\ \ea
\label{parameters1}
\en
Here $\theta_1$ ($\theta_2$) is the angle between the $P_1$ ($P_2$)
three-momentum vector in the $V_1 ( V_2)$ rest frame and the $V_1$
($V_2$) three-momentum vector defined in the $B$ rest frame, and
$\phi$ is the angle between the normals to the planes defined by
$P_1 P_1^\prime$ and $P_2 P_2^\prime$, in the $B$ rest frame.

To take the advantage of more easily extracting the CP-odd and
-even components, the angular distribution is often written in the
linear polarization basis, which is defined, according to the
notation in \cite{transversity}, in the following form
of the decay amplitude:
\beq\label{ampl} A(B_q(t) \to V_1V_2) = \frac{A_0(t)}{x}
{\epsilon}^{*L}_{V_1} {\epsilon}^{*L}_{V_2} - A_{\|}(t)
{\epsilon}^{*T}_{V_1} \cdot {\epsilon}^{*T}_{V_2} / \sqrt{2}  - i
A_{\perp}(t) {\epsilon}^*_{V_1} \times {\epsilon}^*_{V_2} \cdot
\hat{\bf p}_{V_2} / \sqrt{2}~, \eeq where $x\equiv p_{V_1}\cdot
p_{V_2}/(m_{V_1} m_{V_2})$ and $\hat{\bf p}_{V_2}$ is the unit
vector along the direction of motion of $V_2$ in the rest frame of
$V_1$. The transversity amplitudes $A_\|$, $A_0$ and $A_\perp$ are
related to the helicity ones by $A_0 = H_0,  A_{\|} =
\frac{1}{\sqrt{2}} \left( H_{+1} + H_{-1} \right)$, and $A_{\perp}
= \frac{1}{\sqrt{2}} \left( H_{+1} - H_{-1} \right)$. The
normalized differential decay rate in terms of transversity
amplitudes is then given by
\begin{eqnarray}
\frac{1}{\Gamma} \frac{d^3\Gamma}{d\cos\psi d\cos\theta d\varphi}
& = & {\frac{9}{8 \pi}} \biggl \{
  \frac{\Gamma_L}{\Gamma} \cos^2 \psi \sin^2 \theta \cos^2 \varphi
+ \frac{\Gamma_{\perp}}{2 \Gamma} \sin^2 \psi \sin^2 \theta \sin^2
     \varphi \nonumber \\
&& + \frac{\Gamma_{||}}{2 \Gamma} \sin^2 \psi \cos^2 \theta
   - \frac{\zeta}{2 \sqrt{2}} \sin 2\psi \sin^2 \theta \sin 2\varphi \\
&& - \frac{\xi_1}{2} \sin^2 \psi \sin 2\theta \sin \varphi
   + \frac{\xi_2}{2 \sqrt{2}} \sin 2\psi \sin 2\theta \cos \varphi
 \biggr \}, \nonumber \label{dgamma}
\end{eqnarray}
where \be\ba{llllll} \frac{\Gamma_L}{\Gamma} & = & \frac{\vert
A_0\vert ^2}{\vert A_0\vert ^2 + \vert A_{||}\vert ^2
   + \vert A_{\perp}\vert ^2}, &
\frac{\Gamma_{\perp}}{\Gamma} & = & \frac{\vert A_{\perp}\vert
^2}{\vert A_0\vert ^2 + \vert A_{||}\vert ^2
   + \vert A_{\perp}\vert ^2}, \\[5mm]
\frac{\Gamma_{||}}{\Gamma} & = & \frac{\vert A_{||}\vert ^2}{\vert
A_0\vert ^2 + \vert A_{||}\vert ^2
   + \vert A_{\perp}\vert ^2}, &
\zeta & = & \frac{Re\left( A_{||}{A_0}^\ast \right)} {\vert
A_0\vert ^2 + \vert A_{||}\vert ^2 + \vert A_{\perp}\vert ^2},
\\[5mm]
\xi_1 & = & \frac{Im\left( A_{\perp} {A_{||}}^\ast\right)} {\vert
A_0\vert ^2 + \vert A_{||}\vert ^2 + \vert A_{\perp}\vert ^2}, &
\xi_2 & = & \frac{Im\left( A_{\perp} {A_0}^\ast \right)} {\vert
A_0\vert ^2 + \vert A_{||}\vert ^2 + \vert A_{\perp}\vert ^2}, \\
\ea \label{parameters}
\en
and we take the rest frame of $V_1$, $V_2$ moves in the x
direction, and the $z$ axis is perpendicular to the decay plane of
$V_2 \to P_2 P_2^{\prime}$ and we assume that ${p_y}(P_2)$ is
nonnegative. $(\theta,\varphi)$ is the angular coordinates of
$P_1$ and $\psi$ is that of $P_2$, both in the rest frame of
$V_1$.

By measuring the six coefficients in the angular distribution of $
B \to VV$ and their corresponding conjugate processes, we can
construct rich CP violating observables, in addition to the usual
partial rate difference $\Delta=\frac{\Gamma- \ov \Gamma}{\Gamma+
\ov \Gamma}$. Since it is easy to extract the CP information from
the measurements in the transversity basis, we will only
concentrate on this basis. With $\eta, \xi_1 ,\xi_2$ for $ B_u^-$
decays, and similarly $\ov \eta$,~~$\ov \xi_1$, $\ov \xi_2$ for $
B_u^+$ decays, the CP violating observables for the transversity
amplitudes can be constructed as: $T_1=\xi_1+{\overline \xi_1}$,
$T_2=\zeta-{\overline \zeta}$, and $T_3=\xi_2+{\overline \xi_2}$
for the processes with the same branching ratios in both $B_u^-$
decays and their conjugate processes. For the processes with
different branching ratios, we could use the same definition for
the unnormalized distributions.

\medskip
{\bf 4.}~~In this letter, we calculate all the angular
distributions and  direct CP violation in the helicity
\footnote{To save the space, we shall only show the results in the
tranversity basis.} and tranversity bases.  Our results
 \footnote{The relevant formulas of decay ampiltudes can be
 found in \cite{CCTK99}. }
   are shown in Table I,
 from which we
find
: (1) all the charmless $B_u \to VV$ decay modes are dominated by
the longitudinal polarized state and the P-wave amplitudes in
these decay modes are small, thus all the charmless $B_u$ decays
are dominated by the CP-even components and have only small
angular correlation asymmetries ({\it i.e.} CP violation in the
angular correlation coefficients) associated with the imaginary
terms, and (2) the imaginary terms appearing in the angular
correlation are all small, especially they  vanish for the $\rho^-
\rho^0$, $ \ov K^{*0} \rho^- , K^{*-} \phi, K^{*-} K^{*0}$ and $
\rho^- \phi$ modes, and the normalized angular correlation
coefficients in those decay modes are the same for the $B_u$ decay
modes and their conjugate modes. The later feature is a general
phenomenon of all
 the processes involving only one factorized amplitude $X^{(B^- V,V)}$
 (defined in Eqs.~(\ref{term})), for example the
$ \ov K^{*0} \rho^- , K^{*-} \phi, K^{*-} K^{*0}$ and $ \rho^-
\phi$ modes in charmless $B_u$ decays. It results from the fact
that the standard CP-violating weak phase and the CP-conserving
perturbative strong phase are all factored out into a common
factor in the processes with only one $X^{(B^- V,V)}$. Thus all
the angular correlation coefficients, which appear in the
distributions in a bilinear from $A_f A^*_g$, are all real and
their differences only show up in the partial rate difference
. One origin of this general feature
comes form the universality ansatz for the nonfactorizable
contributions. A measurement of non-negligible imaginary terms for
these decay modes does indicate a possible deviation of the
universality ansatz and/or a nontrivial phase among  different
amplitudes.

 The factorized amplitude of $B^- \to \rho^- \rho^0$  is
\be
\label{rhorho}
 A(B^-\to\rho \rho^{-})  = {G_F \over \sqrt{2}} \Bigl[ V_{ub} V^{*}_{ud}
(a_1+a_2) - V_{tb} V^{*}_{td}{3\over 2}(a_7+a_9+a_{10})
 \Bigr]X^{(B^- \rho^0,\rho^-)},
\en
where the factorized term $X^{(BV_1,V_2)}$  has the expression:
\be
\label{term} X^{( BV_1,V_2)} &\equiv & \la V_2 | (\bar{q}_2
q_3)_\vma|0\ra\la V_1|(\bar{q}_1b)_\vma|\ov B \ra =-
if_{V_2}m_2\Bigg[ (\vp^*_1\cdot\vp^*_2) (m_{B}+m_{1})A_1^{
BV_1}(m_{2}^2)  \non \\ &-& (\vp^*_1\cdot p_{_{B}})(\vp^*_2 \cdot
p_{_{B}}){2A_2^{ BV_1}(m_{2}^2)\over (m_{B}+m_{1}) } +
i\epsilon_{\mu\nu\alpha\beta}\vp^{*\mu}_2\vp^{*\nu}_1p^\alpha_{_{B}}
p^\beta_1\,{2V^{ BV_1}(m_{2}^2)\over (m_{B}+m_{1}) }\Bigg].
\en
Since the CKM matrix for the tree and penguin contributions are
comparable, this decay is dominated by the largest $a_1$ and hence
it  is  $N_c$-stable. Due to  isospin symmetry, the QCD penguin
does not contribute to this decay mode. The electroweak penguin
(EWP), though making little contributions,  cannot be neglected
when discussing the CP asymmetry: without the EWP contributions,
the partial rate asymmetry will be zero. The small $\Delta$
reflects the small contributions from the EWP and thus the actual
EWP contributions can be determined by the measurement of the
partial rate asymmetry.

 It is instructive to compare  $B^- \to \rho^- \omega$ with $B^- \to
\rho^- \rho$. Although the average branching ratios for these two
decay modes are almost the same, the physics involved are quite
different. The factorized amplitude for $B^- \to \rho^- \omega$ is
\be
\label{rhoomega} A(B^-\to\omega \rho^{-}) &=&
V_{ub}V_{ud}^*\Big\{a_1X^{(B\omega,\rho^{-})}+a_2X_u^{(B\rho^{-},\omega)}\Big\}
\\ &-& V_{tb}V_{td}^*\Big\{ (a_4+a_{10})X^{(B\omega,
\rho^{-})}+ (2a_3+a_4+2a_5+{1\over
2}(a_7+a_9-a_{10})X^{(B\rho^{-},\omega)}\Big\}. \non
\en
Since the CKM factors in the tree and penguin parts are
comparable, this decay mode is still dominated by the tree diagram
with the largest $a_1$. Unlike  the case of $B^- \to \rho^- \rho$,
the QCD penguin does make sizable contributions to $B^- \to \rho^-
\omega$. The QCD penguin contributions, which are smeared out in
the average quantities of the branching ratio, do show their
impacts on  the angular correlation coefficients. With sizable QCD
penguin contributions, direct CP violation in this decay mode can
be as large as $-(5-10) \%$. While the partial rate asymmetry
depends less upon the factorization scheme, the angular
correlation aymmetries are highly sensitive to  $N_c$. Pursuant to
sizable QCD penguin contributions, we wish to emphasize that a
determination of the unitary triangle $\alpha$ via this decay mode
is more promising than via $\rho^- \rho$.

With the replacement  of $\rho^- $  by $K^{*-}$, the QCD penguin
becomes the dominant mechanism in  $B^- \to K^{*-} \omega(\rho)$
due to the CKM factors involved. The comparable tree and penguin
contributions result in  a significant partial rate asymmetry,
however their destructive interference makes the branching ratio
smaller. The general amplitude for $B^- \to K^{*-} \omega$ is
\be
\label{omegaK} A(B^-\to\omega K^{*-}) &=&
V_{ub}V_{us}^*\Big\{a_1X^{(B\omega,K^{*-})}+a_2X_u^{(BK^{*-},\omega)}\Big\}
\\
&-& V_{tb}V_{ts}^*\Big\{ (a_4+a_{10})X^{(B\omega, K^{*-})}+
(2a_3+2a_5+{1\over 2}(a_7+a_9))X^{(BK^{*-},\omega)}\Big\} \non .
\en
While the average branching ratios do not show significant changes
in these three schemes, the predicted CP-violating observables and
imaginary angular correlation asymmetries are dramatically
different. The naive factrization and our preferred factorization
predict a positive partial rate difference, but the large-$N_c$
improved factorization predicts a negative and smaller one. Direct
CP violation in the partial rate asymmetry of this decay mode
in  our preferred factorization scheme (and also in the naive
factorization scheme) can be larger than 40 $\%$. It is also found
that the angular correlation asymmetries, especially  $T_3$,  in
this decay mode and also in the $B^-\to\rho^0 K^{*-}$ decay mode
are not negligible.

The amplitude for $B^-\to\rho^0 K^{*-}$ is
 \be   \label{rhoK}
A(B^-\to\rho^0 K^{*-})&=&
V_{ub}V_{us}^*\Big\{a_1X^{(B\rho^0,K^{*-})}+a_2X_u^{(BK^{*-},\rho^0)}\Big\}
\non \\ &-& V_{tb}V_{ts}^*\Big\{
(a_4+a_{10})X^{(B\rho^0,K^{*-})}+{3\over2}(a_7+a_{9})X^{(B
K^{*-},\rho^0)}\Big\}.
\en
The pattern of $Br(B^-\to\rho^0 K^{*-})>Br(B^-\to\omega K^{*-})$,
which is independent of the factorization scheme, comes from the
behaviour of the QCD penguin: a destructive interference among
 the $a_4$ and $ a_3 + a_5$ terms in $\omega K^{*-}$ makes its branching
ratio smaller than that of $\rho K^{*-}$, where the latter does
not suffer from the destructive interference. Since there is only
one dominant QCD penguin contribution in this $N_c$-stable $\rho
K^{*-}$ mode, the interferences between the tree and penguin
contributions have the same behaviours within three factorization
schemes, and the associated partial rate asymmetries predicted
thus have a definite sign no matter which factorization scheme is
used. Due to $N_c$-insensitivity and the existence of not-so-small
tree contributions, a determination of the unitary triangle $
\gamma$ through this decay mode is encouraging.

All the  analysis and conclusions mentioned above are based on a
positive $\rho$, which is favored by the global fitting. However,
it has been shown \cite{Hou98,CCTK99} that a negative $\rho$ does
show an improvement between the theory and experiments. The impact
of a negative $\rho$ on  $B_u \to V V$ is studied with a simple
sign-flip of $ \rho$ and is  discussed for the following four
different classes. The first class is the $b \to s $
purely-penguin modes, consisting of $\ov K^{*0} \rho^-$ and
$K^{*-} \phi$ , where the CKM factors involved in these decay
modes are not sensitive to $\rho$. Thus the branching ratios for
$B_u^{\pm}$ decays are not sensitive to the sign  of $\rho$ and
almost the same. The partial rate asymmetry is then very small and
not so interesting. The second class is the $b \to d $
purely-penguin modes, consisting of $ K^{*-} K^{*0}$ and $\rho^-
\phi$ , where the involved CKM factor is $V_{tb} V_{td}^*$ which
is very sensitive to the sign of $\rho$. A sign-flip for the
$\rho$ from a positive one to a negative one can enhance the
branching ratio by a factor of two. Because of this enhancement in
the branching ratio, the partial rate difference in these two
decay modes is suppressed. While  $ K^{*-} K^{*0}$ mode is
$N_c$-stable, $\rho^- \phi$ mode is highly sensitive to the
factorization scheme we used, and the associated CP-violating
observables too. While the two classes discussed so far are
purely-penguin processes which are simple to analyze, the next two
classes having both tree and penguin contributions are more
subtle. For the third class, which involves $b \to s $ transition
and consists of $ K^{*-} \rho$ and $ K^{*-} \omega$, the penguin
contribution plays the dominant role with a sizable tree
contribution. The destructive (constructive) interference between
tree and penguin contributions makes the branching ratio smaller
(larger) and partial rate asymmetry larger (smaller) with positive
(negative) $\rho$, respectively. The decays in the last class with
$ \rho^{-} \rho$ and $ \rho^{-} \omega$ have comparable CKM
factors in the tree and penguin parts and thus are dominated by
the largest $a_1$ term. Because of this and the absence of the QCD
penguin contribution, $ \rho^{-} \rho$ has nearly the same
branching ratio as its conjugate process and thus the associated
partial rate difference is small and insensitive to the
factorization schemes and the sign of $\rho$. It is very
interesting that the impact of negative $\rho$ on the $ \rho^{-}
\omega$ mode shows an opposite behaviour to  the third class: one
gets a  smaller (larger)  branching ratio and  larger (smaller)
partial rate difference with a  negative (positive) $\rho$,
respectively.

{\bf 5.}~~In this letter, we revisit two-body charmless
nonleptonic decays of $B \to V V$ by employing  the generalized
factorization approach in which the effective Wilson coefficients
$c^{\rm eff}_i$ are renormalization-scale and -scheme independent
while factorization is applied to the tree-level hadronic matrix
elements. Contrary to previous studies, our $c_i^{\rm eff}$ do not
suffer from gauge and infrared problems. Following the standard
approach, we make a further universal assumption for the
nonfactorizable contributions and thus these nonfactorizable
effects can then be parametrized in terms of $N_c^{\rm eff}(LL)$
and $N_c^{\rm eff}(LR)$, the effective numbers of colors arising
from $(V-A)(V-A)$ and $(V-A)(V+A)$ four-quark operators,
respectively. The full angular distributions of charmless  decays
are calculated in terms of not only  the helicity basis but also
the transversity basis. In addition to the partial rate
difference, we also calculate other CP violating observables.
Results from three different schemes for the nonfactorizable
contribution, the naive factorization, large-$N_c$ improved
factorization and our preferred choice-the optimized
hetero-factorization are presented.

Our main results are the following:

\begin{itemize}
\item  The longitudinal polarization dominates over other polarization
states (about 90$\%$) for all  charmless VV decay modes
 with the beautiful pattern: $A_0>>A_{\|}>A_{\perp}$, {\it i.e.} the
P-wave (CP-odd component) amplitudes are small. Thus all the
charmless $B_u \to VV$ decays are governed by the CP-even
components and the imaginary angular distribution coefficients are
small.
\item For those processes involving only one factorized amplitude
 $X^{(B^- V,V)}$, such as $ \ov K^{*0} \rho^-$ , $K^{*-} \phi$,
$ K^{*-} K^{*0}$, $\rho^- \rho^0 $ and $ \rho^- \phi$ modes, the
imaginary terms appearing in the angular distribution are all
zero. Thus a  measurement of these imaginary terms will show
whether there is a deviation from the universal ansatz for the
nonfactorizable contributions, a possible final state phase, or
even a nontrivial phase from the new physics.
\item Though having the largest branching ratios,
$ B_u^{\pm} \to \rho^{{\pm}} \rho $, which are dominated by the
tree diagrams, have a small partial rate difference. Thus the most
exciting decay mode from the viewpoint of a large branching ratio
and also a large partial rate difference is $ B_u^- \to \rho^-
\omega$, which has a sizable penguin contribution.
Due to this sizable penguin contribution, we would like to
emphasize that a possible determination of the unitary triangle
$\alpha$ is more promising in $ B_u^- \to \rho^- \omega$  than in
$ B_u^- \to \rho^- \rho$.
\item Since the penguin contributions play a major role in
$B_u^- \to K^{*-} \rho(\omega)$, large $N_c$-improved
factorization predicts a large cancellation between the tree $a_1$
and $a_2$ terms and thus a smaller partial rate asymmetry.
 For $B_u^- \to K^{*-} \rho$, which is not sensitive to the information of
 nonfactorization,  the CP-violating
observables  have the same sign in these three factorization
schemes. Because of  the $N_c$-insensitivity and the not-so-small
tree contributions, a determination of the unitary triangle $
\gamma$ through this decay mode is encouraging. Though having the
largest partial rate asymmetry, $B^- \to K^{*-} \omega$ suffers
from the theoretical uncertainty in the nonfactorizable contents.
The quantum interference among the tree and penguin contributions
is changed
and thus the relevant sign of the partial rate asymmetry is also
changed when we use the large-$N_c$ improved factorization instead
of the naive factorization. Direct CP violation in the partial
rate asymmetry of the  $B^- \to K^{*-} \omega$ mode
 in our preferred factorization scheme (and also in the naive
factorization scheme) can be larger than 40 $\%$. The angular
correlation asymmetries predicted for  $B^-\to \rho^0(\omega)
K^{*-}$  are not negligible.
\item To show the parametric correlation with  $\rho$, we also
make some predictions based on a negative $\rho$. A negative
$\rho$ will enhance the penguin contributions with $V_{td}$ which
is proportional to $(1-\rho-i \eta)$ and change the interference
among the tree and penguin contributions especially for those
involved $V_{ub}$ as $(\rho - i \eta)$. A negative $\rho$ has
little impact on the $b \to s$ purely-penguin modes, but can
enhance the branching ratios and thus reduce the associated
partial rate asymmetries of $b \to d$ purely-penguin modes by a
factor of two with a sign-flip done in this analysis. For the
decay modes with both tree and penguin contributions, there are
two totally different behaviours: the branching ratios are
enhanced (reduced) and the partial rate differences are reduced
(enhanced) for the $b \to s$ ($b \to d$) transition, except for
the $ \rho^{-} \rho$ mode which is governed by the tree
contribution and has nearly the same branching ratio as its
conjugate process. The partial rate difference for $ \rho^{-}
\rho$ is small and insensitive to the factorization schemes and
the sign of $\rho$.
\end{itemize}

Finally, we discuss some uncertainties in our calculation and
their possible impacts.
\begin{itemize}
\item In this letter, we have
neglected the $W$-annihilation (WA) and the space-like penguin
(SP) contributions. The WA and SP do not appear in the $B_u^- \to
\rho^- \phi$ mode (likewise, the $W$-exchange contribution and SP
also disappear in  $B_d \to \rho(\omega) \phi$), thus these
processes do not suffer from the uncertainties due to nonspectator
contributions. The impact of WA on the purely-penguin modes may be
quite significant. For $B_u^- \to K^{*-} \phi$, the CKM-suppressed
WA with the largest $a_1$ may have a large effect under the
condition that a large cancellation among the QCD penguin
contributions happens. Likewise for $B_u^- \to K^{*-} K^{*0}$, WA
with the largest $a_1$ may also have a large influence on this
decay mode because of the comparable CKM factors for the penguin
and WA.
\item   The perturbative strong phase is included in our calculation,
while soft final-state interaction (FSI) phases are not considered
in this paper. The soft FSI phases do have large impacts on the
angular correlation coefficients, especially for classes involving
one $X^{(BV,V)}$ where the imaginary terms are all vanishing.
However, a measurement of non-vanishing  imaginary terms for this
class of decay modes is not necessarily claimed to be a direct
confirmation of the FSI phase in B decays since they can be
generated by the CP-violating phases induced from  new physics.
\item  Our
results for  CP observables are evaluated at $k^2=m_b^2/2$. It is
known that  CP violation is sensitive to the $k^2$ we used.
Besides, the physics of $B_u \to VV $ is sensitive to the form
factors we used as shown in the recent paper \cite{CCTK99}. These
topics will be discussed in a separate publication.
\end{itemize}

\bigskip
\noindent ACKNOWLEDGMENTS:~~We wish to thank Prof. H.Y. Cheng and
Prof. L. Wolfenstein
 for valuable discussions. One of us
(B. T.)  is grateful to Prof. R. Enomoto for discussions about
experiments and to Prof. K. Hagiwara and Prof. Y. Okada for useful
discussions. He also thanks the KEK Theory Group for the
hospitality and financial support during his visit. His work is
supported in part by the National Science Council of the Republic
of China under Grant NSC88-2112-M006-013.  C.W. Chiang is indebted
to Prof. Fred Gilman for his continual support and consideration,
and his work is supported in part by the Department of Energy
under Grant No. DE-FG02-91ER40682. He also thanks for the
hospitality of Academia Sinica during his visit.


\renewcommand{\baselinestretch}{1.1}
\newcommand{\bi}{\bibitem}
%

\newpage
\begin{table}
\noindent Table A.1 \quad  Branching ratios, partial rate
asymmetries, angular correlation coefficients in the transversity
basis  for various $B_u^- \to VV$ processes.
Results are obtained  using the BSW form factors in the absence of
soft FSI phases. The Wolfenstein parameters
$(\rho,\eta)=(0.175,0.370)$ are used in different factorization
schemes (FS's): (a) the naive factorization scheme, (b) the
large-$N_c$ improved factorization scheme, and (c) our preferred
factorization scheme: ($(N_{c}(LL),N_{c}(LR))=(2,5)$. The impact
of a negative $\rho$ is studied in (d) our preferred factorization
scheme: ($N_{c}(LL),N_{c}(LR))=(2,5)$ with
$(\rho,\eta)=(-0.175,0.370)$. Values in the parentheses correspond
to the conjugate processes.
\begin{center}
\begin{tabular}{||l|c|c|c|c|c|c|c|c|c||}
\hline \hline Process & FS & BR & $a_{CP}$ &
$\frac{\Gamma_L}{\Gamma}$ & $\frac{\Gamma_{||}}{\Gamma}$ &
$\frac{\Gamma_{\perp}}{\Gamma}$ & $\zeta$
& $\xi_1$
& $\xi_2$
\\
        & & [\tmsix] & [\%] &[\%]  &[\%]  &[\%]  &[\%]  & [\tmthree] &
[\tmthree] \\ \hline
$\pidfour$ & (a) & 4.80 & 0.21 & 84.2 &9.6 & 6.2 & $-$28.5 &  0 &
0 \sc
 & & (4.78) & --- & (84.2) & (9.6) & (6.2) & ($-$28.5) & ( 0)
& ( 0)  \sc
\hline
 & (b) & 0.290 & 0.69  & 84.2 & 9.6 & 6.2 & $-$28.5 &  0 &
 0  \sc
 & & (0.286) & --- & (84.2) & (9.6) & (6.2) & ($-$28.5) & ( 0)
& ( 0)  \sc
\hline
 & (c) & 3.92 & 0.13 & 84.2 & 9.6 & 6.2 & $-$28.5 &  0 &
 0  \sc
 & & (3.91) & --- & (84.2) & (9.6) & (6.2) & ($-$28.5) & ( 0)
& ( 0)  \sc
\hline
 & (d) & 3.79 & 0.13 & 84.2 & 9.6 &6.2 & $-$28.5 & 0 &
 0  \sc
 & & (3.78) & --- & (84.2) & (9.6) & (6.2) & ($-$28.5) & ( 0)
& ( 0)  \sc
\hline
$\pidtwo$ & (a) & 5.150 & 0.05  & 87.6 & 7.1 & 5.3 & $-$24.9 &  0
&  0 \sc
 & & (5.145) & --- & (87.6) & (7.1) & (5.3) & ($-$24.9) & ( 0)
& ( 0)  \sc
\hline
 & (b) & 7.82 &  0.06  & 87.6 & 7.1 & 5.3 & $-$24.9 &  0 &
 0  \sc
 & & (7.81) & --- & (87.6) & (7.1) & (5.3) & ($-$24.9) & ( 0)
& ( 0)  \sc
\hline
 & (c) & 4.03 & 0.12  & 87.6 & 7.1 & 5.3 & $-$24.9 &  0 &
 0  \sc
 & & (4.02) & --- & (87.6) & (7.1) & (5.3) & ($-$24.9) & ( 0)
& ( 0)  \sc
\hline
 & (d) & 3.892 &  0.01  & 87.6 & 7.1 & 5.3 & $-$24.9 &  0 &
 0  \sc
 & & (3.891) & --- & (87.6) & (7.1) & (5.3) & ($-$24.9) & ( 0)
& ( 0)  \sc
\hline

$\pidone$ & (a) & 4.78 & 20.40  & 88.3 & 6.8 & 4.9 & $-$24.5 &
$-$0.06 & 5.45  \sc
 & & (3.16) & ---  & (88.6) & (6.7) & (4.8) & ($-$24.3) & (0.25) &
($-$2.21)  \sc
\hline
 & (b) & 5.76 & 9.48  & 88.3 & 6.8 & 4.9 & $-$24.5 & $-$1.08 &
9.76  \sc
 & & (4.76) & ---  & (88.7) & (6.6) & (4.7) & ($-$24.2) & (0.62) &
($-$5.63)  \sc
\hline
 & (c) & 4.43 & 24.43  & 88.3 & 6.8 & 4.9 & $-$24.5 & $-$0.30 &
3.00  \sc
 & & (2.69) & ---  & (88.5) & (6.7) & (4.8) & ($-$24.3) & ($-$0.03)
& (0.30)  \sc
\hline
 & (d) & 7.68 & 12.69  & 88.2 & 6.8 & 5.0 & $-$24.5 & $-$0.22 &
1.94 \sc
 & & (5.95) & ---  & (88.3) & (6.8) & (4.9) & ($-$24.5) & ($-$0.04) &
(0.39)  \sc
\hline

$\pidthree$ & (a) & 2.56 & 42.62 & 87.4 & 7.2 & 5.4 & $-$25.1 &
0.39 & $-$3.38  \sc
 & & (1.03) & --- & (87.2) & (7.3) & (5.5) & ($-$25.3) & ($-$0.24) &
(2.09)  \sc
\hline
 & (b) & 1.62 & $-$9.70 & 91.2 & 5.5 & 3.3 & $-$22.1 & 5.41 &
$-$46.6  \sc
 & & (1.99) & --- & (91.5) & (5.4) & (3.1) & ($-$22.1) & ($-$2.62)
& (22.6)  \sc
\hline
 & (c) & 2.76 & 46.03 & 87.7 & 7.1 & 5.2 & $-$24.9 & 0.59 &
 $-$5.06
\sc
 & & (1.02) & --- & (87.3) & (7.3) & (5.4) & ($-$25.2) & ($-$0.66)
& (5.65)  \sc
\hline
 & (d) & 4.93 & 21.13 & 87.7 & 7.1 & 5.2 & $-$24.9 & 0.30 &
 $-$2.56
\sc
 & & (3.21) & --- & (87.6) & (7.1) & (5.3) & ($-$25.0) & ($-$0.27)
& (2.33)  \sc
\hline
\hline
\end{tabular}
\end{center}
\end{table}

\newpage
\begin{table}
\noindent Table A.1 \quad
(continued)
\begin{center}
\begin{tabular}{||l|c|c|c|c|c|c|c|c|c||}
\hline \hline Process & FS & BR & $a_{CP}$ &
$\frac{\Gamma_L}{\Gamma}$ & $\frac{\Gamma_{||}}{\Gamma}$ &
$\frac{\Gamma_{\perp}}{\Gamma}$ & $\zeta$
& $\xi_1$
& $\xi_2$
\\
        & & [\tmsix] & [\%] &[\%]  &[\%]  &[\%]  &[\%]  & [\tmthree] &
[\tmthree] \\
 \hline \hline
 $\pidfive$ & (a) & 0.26 & $-$4.97 & 87.5 & 7.6 &
4.9 & $-$25.8 &  0 &  0 \sc
 & & (0.29) & --- & (87.5) & (7.6) & (4.9) & ($-$25.8) & ( 0)
& ( 0)  \sc
\hline
 & (b) & 0.39 & $-$4.37 & 87.5 & 7.6 & 4.9 & $-$25.8 &  0 &
 0  \sc
 & & (0.43) & --- & (87.5) & (7.6) & (4.9) & ($-$25.8) & ( 0)
& ( 0)  \sc
\hline
 & (c) & 0.20 & $-$4.74 & 87.5 & 7.6 & 4.9 & $-$25.8 &  0 &
 0  \sc
 & & (0.22) & --- & (87.5) & (7.6) & (4.9) & ($-$25.8) & ( 0)
& ( 0)  \sc
\hline
 & (d) & 0.36 & $-$2.69 & 87.5 & 7.6 & 4.9 & $-$25.8 &  0 &
 0  \sc
 & & (0.38) & --- & (87.5) & (7.6) & (4.9) & ($-$25.8) & ( 0)
& ( 0)  \sc
\hline
$\pidsix$ & (a) & 0.0136 & 0.37 & 84.3 & 9.0 &6.7 & $-$27.5 &  0 &
0 \sc
 & & (0.0135) & --- & (84.3) & (9.0) & (6.7) & ($-$27.5) & (
0) & ( 0)  \sc
\hline
 & (b) & 0.27 & $-$3.03 & 84.3 & 9.0 & 6.7 & $-$27.5 &  0 &
 0 \sc
 & & (0.29) & --- & (84.3) & (9.0) & (6.7) & ($-$27.5) & ( 0)
& ( 0)  \sc
\hline
 & (c) & 0.0096 & $-$1.11 & 84.3 & 9.0 & 6.7 & $-$27.5 &  0
&  0 \sc
 & & (0.0094) & --- & (84.3) & (9.0) & (6.7) & ($-$27.5) & ( 0)
& ( 0)  \sc
\hline
 & (d) & 0.0175 & $-$0.57 & 84.3 & 9.0 & 6.7 & $-$27.5 &  0 &
 0 \sc
 & & (0.0173) & --- & (84.3) & (9.0) & (6.7) & ($-$27.5) & ( 0)
& ( 0)  \sc
\hline
$\pidseven$ & (a) & 13.53 & $-$0.40 & 90.5 & 5.4 & 4.1 & $-$22.1 &
 0 &  0 \sc
 & & (13.64) & --- & (90.5) & (5.4) & (4.1) & ($-$22.1) & ( 0)
& ( 0)  \sc
\hline
 & (b) & 7.61 & $-$0.46 & 90.5 & 5.4 & 4.1 & $-$22.1 &  0 &
 0  \sc
 & & (7.68) & --- & (90.5) & (5.4) & (4.1) & ($-$22.1) & ( 0) &
( 0)  \sc
\hline
 & (c) & 17.13 & $-$0.38 & 90.5 &5.4 & 4.1 & $-$22.1 &  0 &
 0  \sc
 & & (17.26) & --- & (90.5) & (5.4) & (4.1) & ($-$22.1) & ( 0)
& ( 0)  \sc
\hline
 & (d) & 16.17 & $-$0.37 &90.5 & 5.4 & 4.1 & $-$22.1 &  0 &
 0  \sc
 & & (16.29) & --- & (90.5) & (5.4) & (4.1) & ($-$22.1) & ( 0)
& ( 0)  \sc
\hline
$\pideight$ & (a) & 12.75 & $-$9.51 & 90.5 & 5.4 & 4.1 & $-$22.2 &
$-$0.014 & 0.23  
\sc
 & & (15.43) & --- & (90.5) & (5.4) & (4.1) & ($-$22.2) & (0.012) &
($-$0.19)  \sc
\hline
 & (b) & 7.98 & $-$5.17 &90.5 & 5.4 & 4.1 & $-$22.2 & 0.0040 &
$-$0.066  \sc
 & & (8.85) & --- & (90.5) & (5.4) & (4.1) & ($-$22.2) & ($-$0.0039)
& (0.061)  \sc
\hline
 & (c) & 15.69 & $-$8.43 & 90.5 & 5.4 & 4.1 & $-$22.2 & $-$0.008 &
0.12  \sc
 & & (18.58) & --- & (90.5) & (5.4) & (4.1) & ($-$22.2) & (0.007) &
($-$0.10)  \sc
\hline
 & (d) & 12.03 & $-$10.72 & 90.5 & 5.4 & 4.1 & $-$22.2 & $-$0.006 &
0.10  \sc
 & & (14.92) & --- & (90.5) & (5.4) & (4.1) & ($-$22.2) & (0.011) &
($-$0.18)  \sc
\hline
\hline
\end{tabular}
\end{center}
\end{table}

\end{document}